\documentstyle[prb,aps,multicol,epsf]{revtex}

\begin{document}

\title{Polyelectrolyte adsorption}

\author{David Andelman$^{\dagger}$ and Jean-Fran\c{c}ois Joanny$^{*}$}
\address{$^{\dagger}$School of Physics and Astronomy,
Raymond and Beverly Sackler Faculty of Exact Sciences\\
Tel Aviv University, 69978 Ramat Aviv, Israel\\
$^{*}$Institut Charles Sadron, 6 rue Boussingault, 67083
Strasbourg Cedex, France}

\date{November 2, 2000}
\maketitle

\begin{abstract}
The problem of charged polymer chains (polyelectrolytes) as they
adsorb on a planar surface is addressed theoretically. We review
the basic mechanisms and theory underlying polyelectrolyte
adsorption on a single surface in two situations: adsorption of a
single charged chain, and adsorption from a bulk solution in
$\theta$ solvent conditions. The behavior of flexible and
semi-rigid chains is discussed separately and is expressed as
function of the polymer and surface charges, ionic strength of
the solution and polymer bulk concentration. We mainly review
mean-field results and briefly comment about fluctuation
effects. The phenomenon of polyelectrolyte adsorption on a planar
surface as presented here is of relevance to the stabilization of
colloidal suspensions. In this respect we also mention
calculations of the inter-plate force between two planar surfaces
in presence of polyelectrolyte. Finally, we comment on the problem
of charge overcompensation and its implication to multi-layers
formation of alternating positive and negative polyelectrolytes on
planar surfaces and colloidal particles.

\bigskip

Nous discutons les approches th\'eoriques
qui permettent d'\'etudier les polym\`eres
charg\'es qui s'adsorbent sur une surface plane dans deux situations:
l'adsorption d'un polym\`ere isol\'e et l'adsorption \`a partir d'une solution
lorsque le solvant est un solvant $\theta$. Nous
consid\'erons s\'epar\'ement le
comportement de polym\`eres rigides et flexibles en fonction de divers param\`etres
physicochimiques comme la charge du polym\`ere, la charge de la surface, la
force
ionique de la solution et la concentration en polym\`ere. Nous d\'ecrivons
essentiellement les mod\`eles de type champ moyen mais nous discutons aussi
bri\`evement
l'effet des fluctuations. L'adsorption de poly\'electrolytes est capitale pour
l'\'etude de la stabilit\'e de
suspensions colloïdales. Dans cette optique, nous mentionnons aussi des \'etudes
sur les forces entre deux surfaces planes en pr\'esence de poly\'electrolytes.
Enfin nous pr\'esentons le probl\`eme de l'inversion de charge et son
application \`a
la formation de multicouches poly\'electrolytes sur des surfaces planes et des
particules colloidales.
\end{abstract}


\begin{multicols}{2}

\section{Introduction}

Polymers are often used as chemical additives to control the
stability or tune the viscoelastic properties of colloidal
suspensions. In these applications the interaction between polymers and colloidal
particles in solution plays a major role.
Neutral polymers are mostly used in organic solvents. Their
interaction with colloidal particles has been studied in great
details, both theoretically and experimentally, and their
adsorption behavior is rather well understood \cite{napper,netzdavid,johner}.
Because polymer additives are often charged in aqueous solvents,
it is of great importance to understand
quantitatively their interaction with colloidal
particles.

In this short review we try to give an overview of the
adsorption behavior of charged polymers (polyelectrolytes)
on colloidal particles.
We focus on the limit where the polymer size is much smaller than
the particle size. Hence, it is sufficient to study the adsorption of
polyelectrolyte chains on flat surfaces. Although no complete
description of polyelectrolyte adsorption exists at present, some
important results have recently been obtained  using mean-field
theory and other approximations. These theoretical results seem to
be in good agreement with the experimental observations. One of
the points that we would like to highlight is the possibility of
surface charge inversion. A charged polymer adsorbing on an
oppositely charged surface can overcompensate the original
surface charge. Because the polyelectrolyte creates a thin
adsorbing layer adjacent to the surface, the overall charge of
the surface combined with the adsorbing polymer can have an
opposite sign than the original surface charge.

Let us start by considering single chain behavior in a dilute
aqueous solution of polyelectrolytes \cite{barrat1}.
At this low
concentration limit, the polyelectrolyte chains only weakly
interact with one another and can be considered as isolated chains
dispersed in solution. For simplicity we consider as a
reference only neutral polymers in a $\theta$ solvent,
where the chain conformation obeys Gaussian statistics. The
typical radius of a neutral chain of $N$ monomers is
$R_{0}^2 =2 N a l_{0}$
where $a$ is the monomer size and $l_{0}$ the persistence length.
Charged chains are highly stretched as compared to neutral chains
with the same chemical composition but with no dissociated
charged groups.
 The repulsive interaction energy
between two charged monomers at a distance $r$ can be written as
$v(r)=e^2/4 \pi \varepsilon r=k_{\rm B}T{l_{\rm B}}/{r}$ where $l_{\rm
B}=e^{2}/4\pi \varepsilon k_{\rm B}T$ is the Bjerrum length,
$\varepsilon$ is the dielectric constant of the aqueous medium,
and $k_{\rm B}T$ the thermal energy. For water with
$\varepsilon\simeq 80 \varepsilon_0$ and at room
temperature, $\l_{\rm B}\simeq
7$\AA.
When salt (electrolyte)
is added to the solution, the electrostatic interaction is
screened over the Debye length $\kappa^{-1}=(8\pi nl_{\rm
B})^{1/2}$ where $n$ is the salt concentration.

Let us now consider
two types of polyelectrolytes: flexible and semi-rigid  chains.
If the polymer backbone is semi-rigid or if the polymer is highly charged,
the main effect of the electrostatic interaction is to increase
the persistence length \cite{odijk1}.
The effective persistence length due to electrostatic
interaction is $l_{\tau}=l_{0}+({\tau ^{2}l_{\rm B}})/({4\kappa
^{2}})$ where $\tau =f/a$ is the charge density per unit length
and $0 \le f \le 1$ is the fraction of charged monomers on the
chain.
In the limit of low ionic strength the polymer
becomes very stiff and behaves as a rigid rod ($R=Na$).

If the polymer backbone is flexible, or the polymer is only
weakly charged, the statistics of polymer conformation is
Gaussian at small length scales but the chain is stretched at
larger length scales. A simple way to describe the chain
conformation is shown in Fig. 1 and uses
the so-called ``electrostatic blob" model \cite{pgg1}.
The chain can be viewed as a fully stretched chain of
Gaussian subunits (called the electrostatic blobs) of size
$\xi_{\rm el}\simeq a(a/f^{2}l_{\rm B})^{1/3}$ and each
containing $g$ monomers. Within each blob the chain segment obeys
Gaussian statistics, $g\simeq (\xi_{\rm el} /a)^{2}$. The blob
size $\xi_{\rm el}$ is obtained by equating the electrostatic
interaction of $fg$ charged monomers inside the blob with the
thermal energy $k_{\rm B}T$. The size of the entire chain is then
$R\simeq Na^{2/3}(f^{2}l_{\rm B})^{1/3}$. The crossover between
the rigid and the flexible polyelectrolyte behaviors occurs when
$a^{2}>f^{2}l_{\rm B}l_{0}$.

In the remaining of the paper we review the
adsorption of flexible and semi-rigid
polyelectrolyte chains on planar surfaces. We
discuss the adsorption of a single polyelectrolyte chain on a
charged surface of opposite charge. Section 3 is devoted to the
mean-field theory of multi-chain polyelectrolyte adsorption on a
charged surface. We first summarize the results obtained
analytically by asymptotic methods and then present
some scaling arguments based on
numerical studies of the mean-field equations. In section 4, we
comment on  correlation effects of semi-rigid (rod-like) and
charged polymers. The
last section is devoted to two specific examples: (i) interactions
between surfaces coated by an adsorbed polyelectrolyte layer, and
(ii) polyelectrolyte multilayers.

\section{Single chain adsorption}

In this section we discuss the adsorption of an isolated polymer
chain on an oppositely charged surface.  First we consider
flexible and weakly charged chains and then semi-rigid chains.

\subsection{Flexible and weakly charged chains}

When a flexible polyelectrolyte is attracted by an oppositely
charged surface with surface charge density (in units of number
per unit area)
$\sigma$, the polymer adsorbs and forms a thin layer of thickness
$\delta$ adjacent to the surface. This is shown
schematically on Fig. 2. The chain thickness results
from a balance between the attractive electrostatic interaction
and the repulsive confinement force due to the Gaussian chain
entropic elasticity \cite{borisov}.
The  thickness resulting from the minimization
of the adsorbed chain free energy is:

\begin{equation}
\label{delta} \delta =(a^{2}/f\sigma l_{\rm B})^{1/3}
\end{equation}

This result is valid at sufficiently
large $\sigma$ where the thickness $\delta$ is smaller than the
Gaussian radius $R_0$. At lower values of $\sigma$, the
chain of blobs retains its conformation
upon adsorption and is only weakly adsorbed (see Ref.~\cite{borisov}
for more details). Another polymer-surface
interaction originates from the dielectric discontinuity  between
the surface and the aqueous solution. The dielectric
discontinuity can be understood in terms of repulsive image
charges that cause repulsive interaction at surfaces having a
dielectric constant lower than that of the bulk solution. This
interaction competes with the adsorption mechanism but is rather
weak for moderately high $\sigma$ and
can be ignored in most situations.

When salt is added to the aqueous solution the electrostatic
interaction close to the surface is screened. The conformation
of the adsorbed chain does not change as long as the Debye
screening length  is larger than the thickness, $\kappa ^{-1}>
\delta$. If the screening length becomes shorter than $\delta$,
the chain no longer adsorbs \cite{wiegel}.
This gives the criterion for
adsorption: $\sigma>\kappa ^{3}a^{2}/(f\sigma l_{\rm B})$. The
last result can be understood also by considering the screening
effect on the free energy of the adsorbing chain.

\subsection{Semi-rigid chains}

Using similar arguments it is possible to describe the adsorption
of semi-rigid chains on planar surfaces. The force that opposes
confinement does not originate from the Gaussian chain elasticity
but is due to bending fluctuations of semi-rigid chains
\cite{odijk}. The thickness of the adsorbed chain is

\begin{equation}
\label{deltar} \delta =(l_{\rm B}\sigma \tau l_{\rm
eff}^{1/3})^{-3/5}
\end{equation}

 The effective persistence
length of the chain $l_{\rm eff}$ includes both the electrostatic
and the bare bending contributions. The electrostatic contribution
to the persistence length does not reduce here to the
contribution calculated by Odijk \cite{odijk} because the length
scales involved in the adsorption can be shorter than the
screening length. This is the reason why we introduce here an
effective persistence length \cite{barrat1}. This has been
discussed in detail in Ref.~\cite{netz1}. When image charge
effects are taken into account, the relevant persistence length
is always the bare persistence length $l_{0}$. When salt is
added, a semi-rigid chain also desorbs when the screening length
is of the order of the chain thickness i.e., if $\sigma <\kappa
^{5/3}/(\tau l_{\rm eff}^{1/3}l_{\rm B})$.

Although it is difficult to perform experiments for very dilute
polyelectrolyte solutions in the bulk, single chain manipulations
are possible at surfaces.  Conformations of individual adsorbed
chains can be probed by pulling a single chain off the surface
with the tip of an atomic force microscope (AFM). These
experiments measure the interplay between the chain elasticity
and the surface-chain electrostatic interaction \cite{chatellier}.
The signature of the electrostatic interaction is an increase of the force with
the distance from the surface up to distances
comparable to the Debye screening length. For larger distances the force profile
reaches a plateau.

\subsection{Counterion release}

When a strongly charged polyelectrolyte adsorbs on an oppositely
charged surface, the counterions that neutralize the chain
charges play an important role. Above a critical charge per unit
length, $\tau_{c}=1/l_{\rm B}$, some of the counterions condense
on the polymer and the polyelectrolyte together with the
condensed counterions has an effective charge (per unit length)
$\tau_{\rm eff}$ lower than its nominal $\tau$ value and
approximately equal to the critical condensation charge
$\tau_{\rm eff}\simeq \tau_c=1/l_{\rm B}$.
This phenomenon is known as counterion or
Manning condensation \cite{manning}.

When the chain is adsorbing on an oppositely charged surface, the
condensed counterions can be released into the solution, because
their release leads to a big increase in counterion translational
entropy. The corresponding free energy gain can be as large as
the direct electrostatic interaction and, thus, is a major
driving force for the adsorption. For a semi-rigid
polyelectrolytes adsorbing on a flat surface, counterion release
has recently been studied \cite{sens}. The main result is that a
complete counterion release occurs only for very large surface
charges.

\section{Mean-field theory of polyelectrolyte adsorption}
\subsection{The adsorption profile equation}

 The simplest approach to study the adsorption of a
multi-chain polyelectrolyte solution on a solid surface is to use
a mean-field theory in analogy to what has been done for the
adsorption of neutral polymers \cite{pgg2}. The polymer conformation is
calculated in a mean-field potential due to the interactions with
the surface and the other polymer chains, $U(z)$, and is
described by $\phi$, the so-called order parameter. This order
parameter is related to the local concentration $c(z)$ at a distance $z$
from the surface by $\phi (z)=\sqrt{c(z)}$. Next, for long enough
chain, $N\gg 1$, it is possible to employ the ground state
dominant approximation. Corrections beyond the ground state
dominance are important if the chain tails have important
contribution to the adsorption energetics. These corrections are
not important when the adsorbing layer is thin and dense, as is the case
for
polyelectrolytes. The equation governing the order parameter is
the standard Edwards equation

\begin{equation}
\label{edwards} -\frac{a^{2}}{6}\frac{d^{2}\phi }{dz^{2}}+(U(z)+E
)\phi =0
\end{equation}

\noindent
where $-E$ is the ground state free energy and can be calculated
by equating the polymer chemical potential in the adsorbed layer
and in the bulk. The mean-field potential $U$ includes both the
local electrostatic potential and the excluded volume interaction
with the other monomers. For simplicity, we consider here only the
case of a $\theta$ solvent where the 2nd virial coefficient
exactly vanishes $v=0$, and write $U(z)=-f\psi(z) +
w^{2}c^{2}(z)/2$ where $w^2$ is the 3rd virial coefficient for
the (repulsive) interactions between monomers, $c=\phi^2$ is the
monomer concentration and $\psi$ the rescaled electrostatic
potential (in units of $e/k_{\rm B}T$). We choose the zero of
electrostatic potential so that far from the surface, $\psi=0$,
and discuss the case where the polymer is negatively charged and
the surface is positive. The electrostatic potential $\psi$
satisfies the Poisson-Boltzmann equation
\begin{equation}
\label{PB} \frac{d^{2}\psi }{dz^{2}}=\kappa ^{2}\sinh \psi +4\pi
l_{\rm B}f\phi ^{2}
\end{equation}

The above equation is the Poisson equation relating the Laplacian
of the electric potential to the total charge in the system, which
is the sum of the small ionic charges (satisfying the Boltzmann
distribution) and the  polymer charges. In addition, we need to
specify the boundary conditions for $\psi$ and $\phi$. The
electrostatic boundary condition is ${d\psi }/{dz}|_0=-4\pi
l_{\rm B}\sigma$ because the surface charge density is fixed, and
the $\phi$ boundary condition can be expressed in terms of the
short range non-electrostatic interactions between the polymer
and the surface ${\phi^{-1}}{d\phi}/{dz}|_0=-{1}/{d}$. We note
that this boundary condition is often used in problems of polymer
adsorption. It conveniently depends only on one parameter $d$.
The extrapolation length $d$ is inversely proportional to the
short range interaction between the polymer and the surface. It
is positive and of the order of a molecular size if the
polymer-surface interaction is attractive and of order $k_{\rm
B}T$ per monomer. The length $d$ is negative if the chains are
repelled by the surface and it vanishes for an impenetrable
surface (a hard wall) where $\phi(0)=0$. We will also consider an
indifferent surface for which $1/d$ vanishes; it is a surface at
the transition between adsorption and depletion (ignoring
electrostatic effects).

This set of equations (3)-(4) can be used to calculate the mean
field  adsorption profiles. The profile equations have been solved
either numerically \cite{varoqui,borukhov} or
using asymptotic analysis. We  summarize
below the main findings, starting with the case of adsorption on
non-charged surfaces.

\subsection{Adsorption on electrically neutral surfaces}

We first consider an electrically neutral surface in the limit of
very small ionic strength ($\kappa^{-1} \rightarrow 0$). For
example, this can be the case of polyelectrolyte adsorption at
liquid-air or liquid-liquid interfaces. The polymer chains are
attracted to the surface by non-electrostatic interactions as is
parameterized by the length $d$, and form a thin layer on the
surface. Only a small fraction of the counterions are localized
in this layer as can be checked a posteriori. To a good
approximation, the electrostatic problem can be solved by
considering the dense adsorbed polymer layer by itself as a
charged surface with a charge density $\sigma=f\Gamma$, where
$\Gamma$ is the monomer surface coverage (per unit area):

\begin{equation}
\Gamma=\int [c(z)-c_{\rm bulk}] dz
\end{equation}
 The electrostatic potential is obtained by solving the standard
Gouy-Chapman problem for the counterion charge density in
presence of a charged surface. At length scales smaller or
comparable with the adsorbed layer thickness, the electrostatic
potential $\psi =2\log (\kappa \lambda_{\rm GC})/2$ is constant,
where the Gouy-Chapman length is defined as $\lambda_{\rm GC}
=1/(2\pi l_{\rm B}\sigma)=1/(2\pi l_{\rm B}f\Gamma)$.

 Close to the surface, the excluded volume
interaction dominates and the monomer concentration decays as a
power law
\begin{equation}
c(z)\approx 1/(z+d/2)
\end{equation}
The electrostatic potential creates a strong barrier against
adsorption that starves the adsorbed layer and reduces the
adsorption when the charge of the polymer increases. At large
distances (though smaller than $\lambda_{\rm GC}$), the
concentration decays exponentially $c(z)\simeq \exp (-2z/
\delta_{n})$ where the layer thickness is
$\delta_{n}=a/\sqrt{-6f\psi}$. Because $\Gamma \simeq a^{-2}$, the
thickness is indeed much smaller than the Gouy-Chapman length and
the counterions are not localized inside the adsorbed polymer
layer in consistent with our assumptions. The strong adsorption
limit requires that $\delta_{n}>d$. This indeed fixes the minimum
strength of the attractive short range potential. Note also that
the adsorption on a charged surface having the same sign as the
polymer charge can be treated in a similar way, but then the
effective charge of the surface is $f\Gamma +\sigma$.

\subsection{Adsorption on oppositely charged surfaces: charge
inversion}

When the polymer adsorbed on a charged surface with an opposite
charge, and in the limit of low ionic strength, the polymer
essentially neutralizes the charge of the surface or even slightly
overcompensates it. The effective charge of the surface coated by
the adsorbed polymer layer is low and, as a first approximation,
the Poisson-Boltzmann equation can be linearized. The polymer
forms a very thin and dense adsorbing layer on the surface almost
free of counterions. An asymptotic expansion of the mean-field
equations can be carried out \cite{joanny1} ignoring the role of the
counterions in the adsorbed layer. If the surface charge is low,
excluded volume interactions are negligible and the thickness of
the adsorbed layer remains comparable with the thickness of a
single adsorbed chain, $\delta$, given by eq. \ref{delta}. The
polymer charge always overcompensates the surface charge by a
small amount, $\Delta \sigma /\sigma \simeq \kappa \delta
(1+\delta^{2}/\xi_{\rm el}^{2})$, where the overcharging is defined as $\Delta
\sigma = f\Gamma - \sigma$.

At higher values of the surface charge, $\sigma >\sigma_{c}\sim
f$, excluded volume becomes important and the thickness
$\delta_{\rm ev}$ of the adsorbed layer increases with the
surface charge $\delta_{ev}\simeq \xi_{\rm el}^{2}/\delta$.
In this regime as well,
the surface charge is weakly overcompensated. In the limit of
high ionic strength, the electrostatic interaction is short range
and essentially contributes to the excluded volume interaction.
The properties of the adsorbed layer strongly depend on the short
range interaction between the polymer and the surface. If the
adsorbing surface is an ideal impenetrable surface (with no other
interactions), there is no adsorption just like the case for a single
chain. If the surface is indifferent, i.e., if the short range
attraction exactly compensates the impenetrable surface repulsion
yielding  $1/d=0$, the overcompensation can be quite large,
$\Delta \sigma =2\sigma$.

Numerical studies of the mean-field equation allow an explicit
calculation of the polymer concentration profiles. In
Ref.~\cite{borukhov} polyelectrolyte adsorption is investigated
using different boundary conditions from the one discussed above.
The results are interpreted in terms of scaling laws. The polymer
concentration was fixed to be zero on the surface, which
effectively means that the adsorption length $d$ vanishes. The
electrostatic boundary condition is chosen to be either a fixed
surface potential $\psi_s=const$ or a fixed  surface charge
$\sigma=const$. The numerical results \cite{borukhov} for
different surface boundaries, different monomer size and polymer
charge indicate that, to a good approximation, the adsorption can
be characterized by a single adsorption length scale, $D$.

For low salt concentration, $D\ll \kappa^{-1}$ and for a constant
$\psi_s$,

\begin{equation}
D\sim f^{-1/2}\psi_s^{-1/2}
\end{equation}
In this limit,
$\psi_s \sim \sigma D l_B$, and these results are
in agreement with the asymptotic analysis discussed above,
$D\sim \delta$.
The polymer
surface excess $\Gamma$ can also be obtained by scaling arguments
and is equal to

\begin{equation}
\Gamma \sim {\psi_s^{3/2}}{f^{-1/2}} \sim {\sigma}{f}^{-1}
\end{equation}
The
asymptotic
analysis predicts in this case a weak charge inversion, which is consistent
with this scaling result. The overcompensation is observed numerically by
calculating the electrostatic potential which is non monotonic. Namely, at some
distance from the surface the electric field is zero: $d\psi/dz=0$.

In high salt conditions, the salt screens all electrostatic
interactions including the repulsive monomer-monomer interactions
and the attractive surface-monomer interactions. The characteristic size of the
adsorbed polyelectrolyte layer
$D$ now depends on $\kappa$,

\begin{equation}
D\sim \frac{\kappa}{f\psi_s}
\end{equation}
Using the appropriate  relation between $\psi_s$ and $\sigma$:
 $\psi_s\sim \sigma l_B /\kappa$, the scaling with $\sigma$
is

\begin{equation}
D \sim \frac{\kappa^2}{f \sigma}
\end{equation}

Because of the Debye screening, the electrostatic interaction
generates an electrostatic contribution to the
2nd virial coefficient
$v_{\rm elec} \sim f^2\kappa^{-2}$ even when the solvent is a $\theta$ solvent.
For large enough values of the polymer charge $f$, the surface
excess is
\begin{equation}
\Gamma \sim \frac{\kappa\psi_s}{f} \sim \frac{\sigma}{f}
\end{equation}
At high salts, the overcompensation is not universal and strongly
depends on the system parameters such as $d$. It is thus
difficult to compare the numerical results to the asymptotic
results presented above in very specific conditions. One of the
ways to cause overcompensation is to change the solvent quality.
For a polymer in a good solvent, the overcompensation increases
with the excluded volume parameter $v$ (the 2nd virial
coefficient). For $v\ge v_{\rm elec}$ a strong overcompensation
is expected. This indeed was obtained in Ref. \cite{borukhov}
from the numerical solution of the profile mean-field equations.

\subsection{Adsorption from dense solutions}

The structure of an adsorbed polyelectrolyte layer strongly
depends on the concentration of the bulk solution. The main
feature of semi-dilute or dense solutions is that the structure
factor $S(q)$ has a strong peak at a wavenumber $q^{*}>0$
indicating  that concentration fluctuations decay in an
oscillatory manner. Any perturbation of the otherwise homogeneous
solution creates a spatial modulation in the polymer
concentration profile with a period $2\pi /q^{*}$. This
phenomenon also manifests itself for adsorbing polyelectrolytes.
Damped oscillations of the concentration profile have been
explicitly calculated in \cite{chatellier2} using a
Debye-H\"uckel (linearized) theory and treating the adsorbing
surface as a perturbation. The modulations exist at weak ionic
strength, while the concentration profile becomes monotonically
decreasing for large ionic strength, ($\kappa
>q^{*}$). It is conceivable to assume that
these damped oscillations occur also beyond the validity of the
linear Debye-H\"uckel regime. They can be related to oscillatory
forces measured in  thin film balance with soap films containing
polyelectrolytes \cite{langevin}.

\section{Adsorption of semi-rigid polyelectrolytes: correlation effects}

The main assumption of mean-field theory is that  only the
average concentration is treated while neglecting concentration
fluctuations. Because electrostatic interactions between adsorbed
chains are very strong, they can lead to pronounced correlations.
In particular, it is important to consider concentration
fluctuations in the direction parallel to an adsorbing surface. A
general treatment of concentration fluctuations is not yet
available for flexible chains. For semi-rigid polymers, a rough
approximation is to assume that semi-rigid polyelectrolytes
adsorbed on a surface of opposite charge form a perfect
two-dimensional smectic phase where the all the rods are parallel
to one of the surface directions and equally spaced in the other
direction \cite{netz1}. A large charge overcompensation is found
under certain ionic strength conditions. The total adsorbed
polymer charge is $f\Gamma =\tau \kappa /\log (\tau
\sigma^{-1}\kappa )$. Charge inversion is thus found at large
enough ionic strength for $\tau \kappa
>\sigma$. This model has been extended to highly charged
polymers taking into account counterion condensation by
Shklovskii and collaborators \cite{nguyen}.

For weakly charged polyelectrolytes, concentration fluctuations
certainly play an important role and the chains adsorbed on
the surface can similarly show a smectic or nematic order. A
model where  the flexible polymers are treated as rods formed by
blobs of chains is
presented in Ref. \cite{dobrynin}. The main result is a strong
charge overcompensation  obtained in the intermediate ionic strength regime.

\section{concluding remarks}

We have described in this short review some of the basic
properties of polyelectrolytes adsorbing on planar surfaces at
thermodynamic equilibrium. Assuming that the system is in
thermodynamic equilibrium  is certainly a strong assumption. The
electrostatic interaction between chains is strong and the
relaxation towards the equilibrium state may be very slow in
certain experiments. The adsorbed polyelectrolyte layer can then
be trapped in a metastable state. A non equilibrium theory taking
into account adsorption kinetics is needed to understand these
experiments. We hope that it will be addressed in future works.

Because polyelectrolyte adsorption is
of great importance in more complex situations, we would like to
briefly discuss two types of experimental setups. In the first the
interactions between surfaces (or large colloidal particles) is
mediated by polyelectrolytes, whereas in the second a multilayered
adsorption of alternating positive and negative polyelectrolytes
is considered.

\subsection{Polyelectrolyte adsorption and inter-colloidal forces}

The interactions between two planar surfaces coated by
polyelectrolytes have been studied using a mean-field approach
and scaling arguments by Borukhov et al. \cite{borukhov1}. Within
mean-field theory, a general contact theorem can be obtained and
leads to an expression for the inter-surface force in this
system. The force is calculated numerically from polyelectrolyte
profiles and analytically, in some limits. The principal result
shows that overcompensation of surface charges by the
polyelectrolyte can lead to effective attraction between two
equally charged surfaces.

At low salt concentration, depicted on Fig.~3, strong repulsion
between the two surfaces results from depletion of the
polyelectrolyte from the inter-surface gap at short separations.
At larger separations when the two polyelectrolyte adsorbing
layers on each of the surface have a strong overlap, the
inter-surface force becomes attractive. The magnitude of this
attraction scales as $f^{1/2}$ and its characteristic length
scale scales as $l_0/f^{1/2}$. At large inter-surface separations
the two surfaces interact only via weak electrostatic repulsion.

At high salt concentration and for highly charged polyelectrolyte
$f\lesssim 1$, the polymer contribution to the attraction at short
distances scales as $f /\kappa$ and the characteristic length
scale is $\kappa l_0^2/f$. An example is shown on Fig.~4. In the
other limit of weakly charged polyelectrolytes, $f\ll 1$, the
interaction is repulsive for all surface separations and decays
exponentially with a decay length equal to $\kappa^{-1}$.

It is worth mentioning that irreversible effects play an important
role on inter-surface forces because of very slow kinetics in the
process of polyelectrolyte adsorption. In many situations it is a
good approximation to consider that the surface excess of adsorbed
polyelectrolytes remains constant as the distance between the two
plates is varied. In Ref. \cite{borukhov1} it was shown that
attractive forces can be obtained for irreversible adsorption and
that their magnitude is larger than the ones corresponding to
equilibrium processes.

\subsection{Polyelectrolyte multilayers}

Polyelectrolyte multilayers are obtained by consecutive
adsorption of positively and negatively polyelectrolytes either on
a flat surface or on a colloidal particle \cite{decher}.
If the process is
initiated from a charged planar surface, the surface charge can be
inverted by adsorption of the first polyelectrolyte layer (of
opposite charge) as discussed in the previous sections. The second
polyelectrolyte layer is bound to the first layer by polyelectrolyte complexation.
A simple Debye-H\"uckel model has recently been proposed in
Ref.~\cite{castelnovo}. This model predicts charge
inversion after adsorption of the second layer, and allows the
built-up of alternating positively and negatively charged
polyelectrolyte layers by successive complexations. The Debye-H\"uckel
approach is a high temperature approximation which ignores strong association
between ions. An alternative approach considering explicitly the formation of
ion pairs has recently been proposed by Zeldovich and Khokhlov
\cite{zeldovich}.

\acknowledgments

One of us (DA) would like to acknowledge financial support from
the Israel Science Foundation founded by the Israel Academy of
Sciences and Humanities --- centers of Excellence Program and the
US--Israel Binational Science Foundation (BSF) under grant no.\
98-00429.



\section*{FIGURE CAPTIONS}

\begin{itemize}

\item
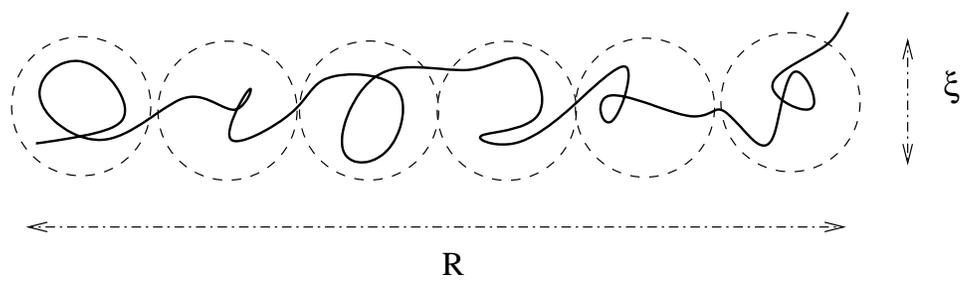
{\bf Fig.~1} ~~~A polyelectrolyte is represented as a rod-like
chain of electrostatic blobs. Inside each blob of size $\xi$, the
chain obeys Gaussian statistics. The chain length is $R$.

\item
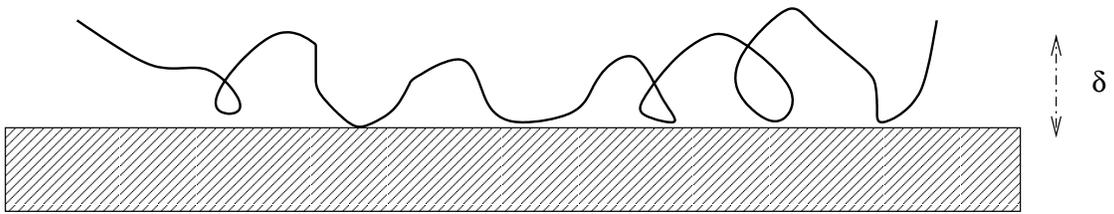
{\bf Fig.~2}~~~A chain adsorbed on a flat solid surface is
schematically drawn. The adsorption length is $\delta$.

\item
{\bf Fig.~3}~~~Inter-surface interactions for a polyelectrolyte
solution between two surfaces held at a constant potential,
$\psi_s=2$, and in low salt condition, $n=10^{-6}$\,M. (a) The
surface free energy $2\pi \Delta F$ as a function of the
inter-surface separation $w$. (b) The force per unit area
$\Pi=-\partial \Delta F/\partial w$ between the two surfaces as a
function of the separation $w$. The different curves correspond
to different polymer charge: $f=1$ (solid curve) and $f=0.2$
(dashed curve). Adapted from Ref.~\cite{borukhov1}.

\item
{\bf Fig.~4}~~~Inter-surface interactions for highly charged
polyelectrolytes ($f=1$) at high salt concentration. (a) The
surface free energy $2\pi \Delta F$ as a function of the
inter-surface separation $w$. (b) The force per unit area
$\Pi=-\partial \Delta F/\partial w$ between the two surfaces as a
function of the separation $w$. The salt concentration is
$n=0.1$\,M (solid curve) and $n=1$\,M (dashed curve). The inset
shows the mid-plane values of the electrostatic potential
$y(0)=\psi(w/2)$ as a function of $w$. Adapted from
Ref.~\cite{borukhov1}.

\end{itemize}

\end{multicols}

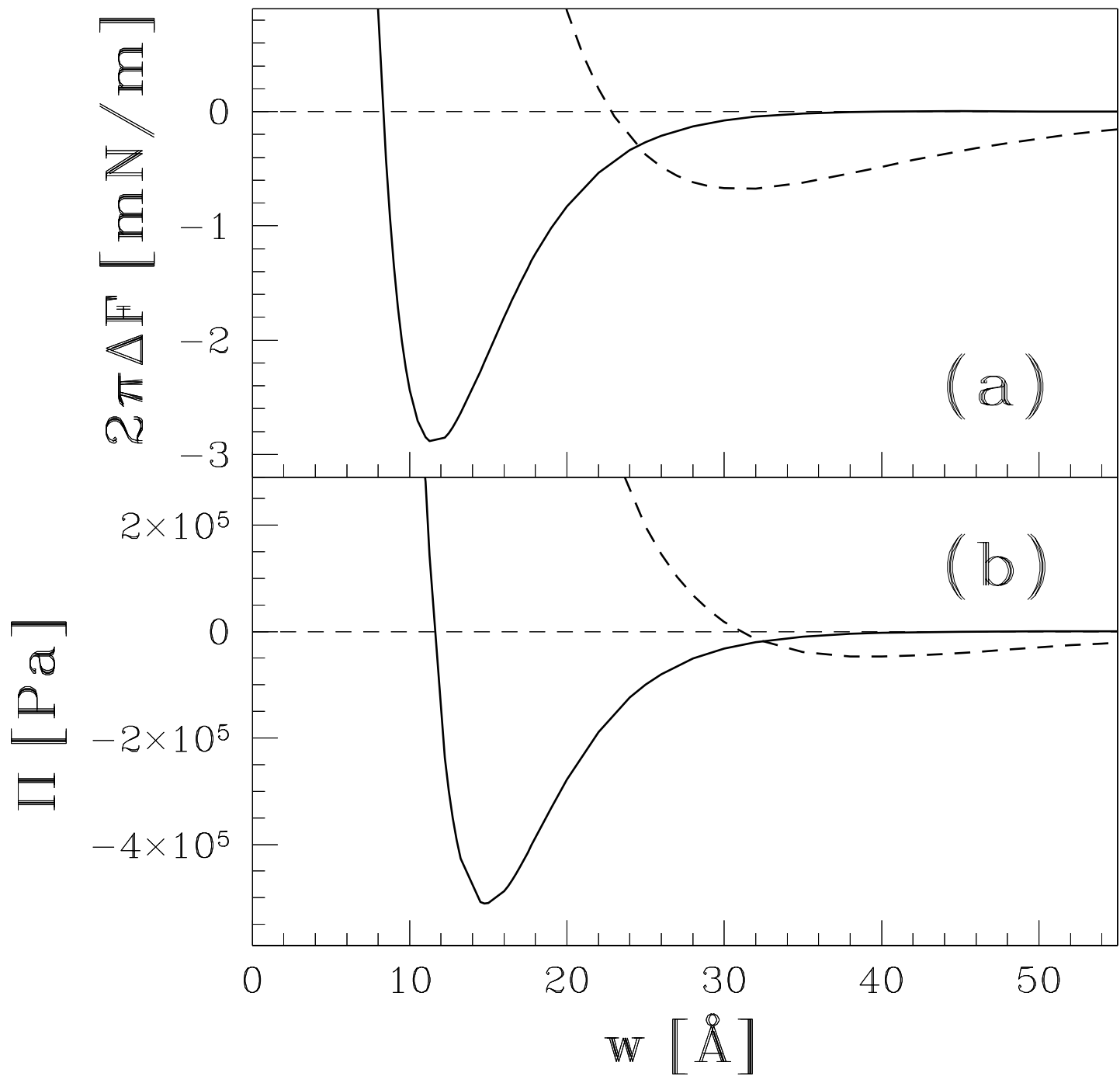
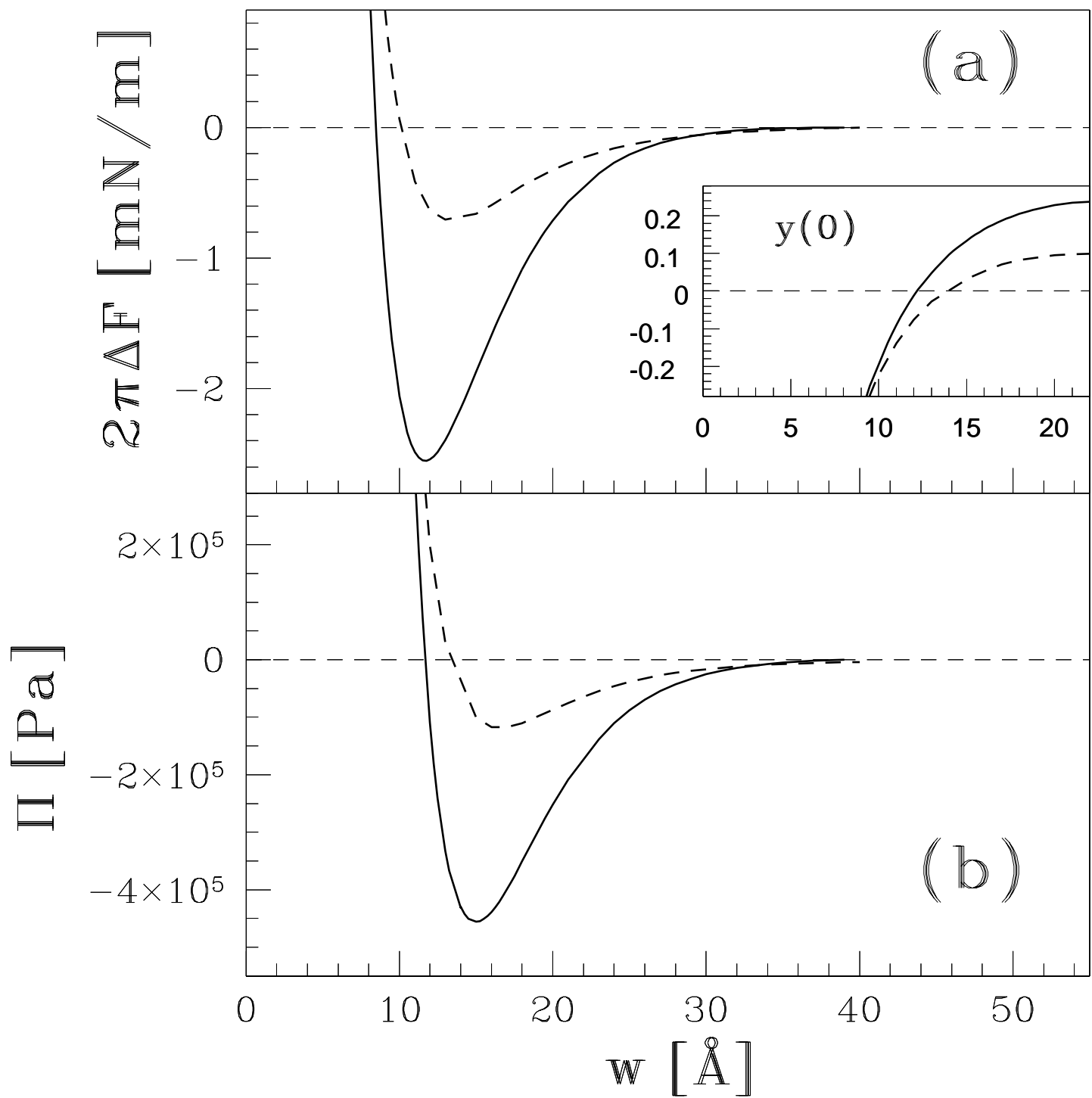
\end{document}